\documentclass[runningheads]{llncs}
\usepackage[T1]{fontenc}
\usepackage{graphicx}
\usepackage{cite}
\usepackage{amssymb}
\usepackage{ bbold }
\usepackage{amsmath}
\usepackage{multirow}
\usepackage{longtable}
\usepackage{hyperref}

\usepackage{tabularx}
\newcolumntype{Z}{>{\centering\arraybackslash}X}
\usepackage{color}

\newcommand{\mycomment}[1]{}
\graphicspath{ {./images/} }

\newcommand{\beginsupplement}{
 \setcounter{table}{0} 
 \renewcommand{\thetable}{S\arabic{table}} 
 \setcounter{figure}{0} 
 \renewcommand{\thefigure}{S\arabic{figure}}
}

\begin{document}

\title{Multi-Dataset Multi-Task Learning for COVID-19 Prognosis}
\author{Filippo Ruffini \inst{1} \and
Lorenzo Tronchin \inst{1} \and Zhuoru Wu \inst{2} \and Wenting Chen \inst{3} \and Paolo Soda \inst{1} \and Linlin Shen \inst{2} \and Valerio Guarrasi \inst{1}\thanks{Corresponding author: Valerio Guarrasi, E-mail: valerio.guarrasi@unicampus.it} }
\authorrunning{F. Ruffini et al.}

\institute{Unit of Computer Systems \& Bioinformatics, Department of Engineering,
Università Campus Bio-Medico di Roma, Rome, Italy \and College of Computer Science and Software Engineering, Shenzhen University, Shenzhen, China \and Department of Electrical Engineering, City University of Hong Kong, Hong Kong, China}
\maketitle 

\begin{abstract}
In the fight against the COVID-19 pandemic, leveraging artificial intelligence to predict disease outcomes from chest radiographic images represents a significant scientific aim. The challenge, however, lies in the scarcity of large, labeled datasets with compatible tasks for training deep learning models without leading to overfitting. Addressing this issue, we introduce a novel multi-dataset multi-task training framework that predicts COVID-19 prognostic outcomes from chest X-rays (CXR) by integrating correlated datasets from disparate sources, distant from conventional multi-task learning approaches, which rely on datasets with multiple and correlated labeling schemes. Our framework hypothesizes that assessing severity scores enhances the model's ability to classify prognostic severity groups, thereby improving its robustness and predictive power. The proposed architecture comprises a deep convolutional network that receives inputs from two publicly available CXR datasets, AIforCOVID for severity prognostic prediction and BRIXIA for severity score assessment, and branches into task-specific fully connected output networks. Moreover, we propose a multi-task loss function, incorporating an indicator function, to exploit multi-dataset integration. The effectiveness and robustness of the proposed approach are demonstrated through significant performance improvements in prognosis classification tasks across 18 different convolutional neural network backbones in different evaluation strategies. This improvement is evident over single-task baselines and standard transfer learning strategies, supported by extensive statistical analysis, showing great application potential.

\keywords{Chest X-rays \and Deep Learning \and CNN \and Transfer Learning}
\end{abstract}


\section{Introduction}

Numerous studies have applied deep learning techniques to COVID-19 for both diagnosis and prognosis. While most focus on image-based AI solutions for COVID-19 diagnosis~\cite{shi2020review}, prognosis prediction receives less attention~\cite{buttia2023covidreviewprognosis}. Many prognostic studies employ single-task learning (STL) methods to predict outcomes like intensive care unit (ICU) admission or mortality~\cite{kulkarni2021prognosis, bae2021predicting, lee2022mortality, li2023progICU, rahman2023mortality}. Alternative approaches predict patient severity progression~\cite{jiao2021prognostication, guarrasi2023multi, schoning2021development, li2022prognostication}, using datasets like AIforCOVID~\cite{soda2021aiforcovid}, to classify patients into mild and severe based on treatment outcomes. Some STL models assess COVID-19 pneumonia severity from CXR using structured severity scores~\cite{zhu2020severity,cohen2020severity, danilov2022severity}, like the Brixia-score~\cite{signoroni2021severity} that segments the lungs into six areas for severity grading~\cite{borghesi2020score}, with these scores statistically linked to treatment outcomes, underlining their prognostic value~\cite{borghesi2022scorecorrelation}. 

Alternately, multi-task learning (MTL) trains models on multiple related tasks simultaneously to enhance reciprocal learning~\cite{caruana1997multitask}. 
In healthcare, studies show that combining two relevant tasks can boost model performance, utilizing various task combinations, architectures for parameter sharing, or loss functions~\cite{zhao2023multi}. Specifically for COVID-19, MTL has been applied to diagnostic predictions and lesion segmentation on imaging data~\cite{amyar2020multi,he2021synergistic}, with some studies exploring the synergy between diagnosis and prognosis for a more integrated disease perspective~\cite{wang2020multi, bao2022multi}. While MTL exhibits promising performance, its primary application remains within datasets having multiple labels per instance. Integrating different data sources for distinct tasks is challenging due to the need for universal annotations across datasets, requiring substantial efforts in data harmonization.

Research in MTL is evolving towards multi-dataset multi-task (MDMT) learning approaches, aiming to leverage diverse data sources to uncover indirect task connections beyond direct data instance links~\cite{xie2023cross-md, kaiser2017one-md, kapidis2021multi-md, hosseini2019distill-md}. This strategy, akin to transfer learning principles, seeks to enrich the learning process by integrating varied datasets. Nonetheless, much of this research does not specifically address medical imaging. For instance, in~\cite{xie2023cross-md} it was implemented an MDMT framework on EEG datasets for action recognition with partially overlapping labels, while in~\cite{kapidis2021multi-md} it was developed a MDMT framework for video-based action recognition, focusing on closely related activities. 

Our MDMT learning model aims to predict COVID-19 severity by integrating two distinct datasets, AIforCOVID~\cite{soda2021aiforcovid} and BRIXIA~\cite{signoroni2021severity}, each highlighting different facets of the same medical condition. BRIXIA focuses on radiological signs of the disease, showcasing the patient's condition shortly after diagnosis, whereas AIforCOVID forecasts future health status based on clinical evolution. By merging these tasks, we intend to show that combining severity assessments can provide deeper insights into patient outcome predictions, thereby enhancing severity classification accuracy. Furthermore, the use of datasets with different labeling schemes offers an innovative approach in healthcare, allowing for efficient multi-source data integration without the need for extensive relabeling, often required in traditional data merging methods. Our main contributions are as follows: \textit{i}) we introduce a MDMT model for two publicly available COVID-19 CXR datasets, each assigned to a different task; \textit{ii}) we implement a multi-task loss function incorporating an indicator function to facilitate multi-task optimization across data sources; \textit{iii}) through an extensive evaluation using 18 CNN architectures, we demonstrate significant performance gains compared to conventional transfer learning and STL approaches.

\section{Methods}
\label{section:method}
\begin{figure}[t!]
\centering
\includegraphics[width=\textwidth]{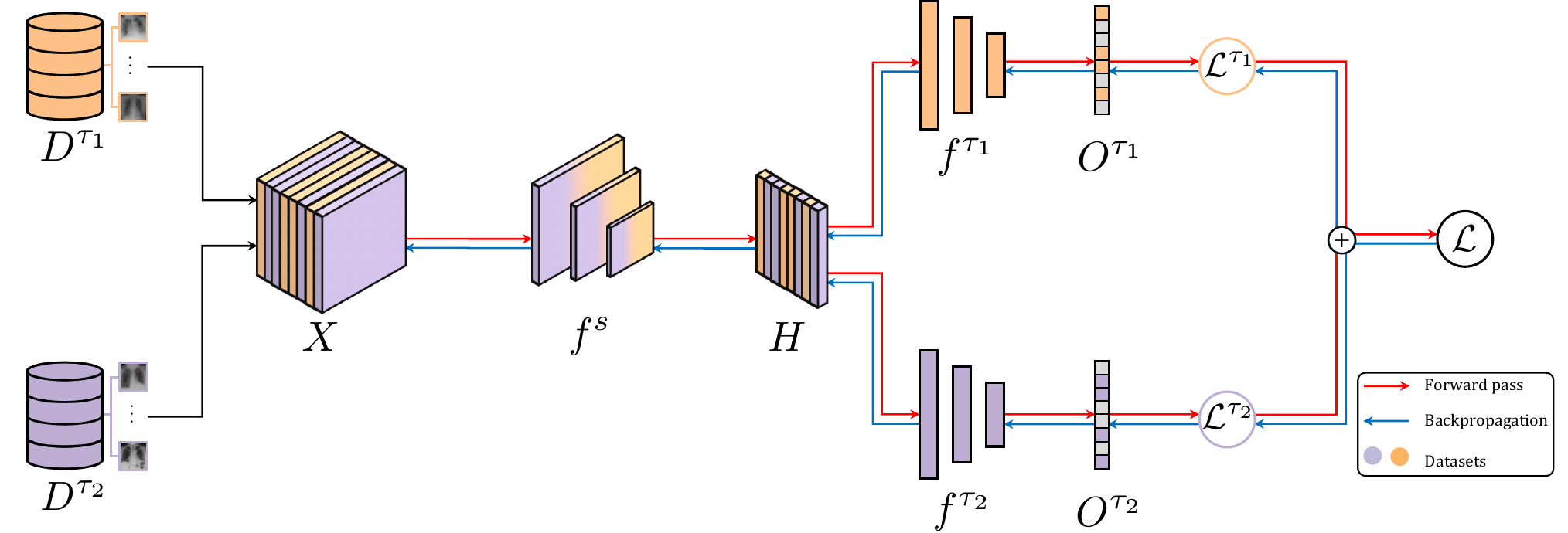}
\caption{Overview of the MDMT model architecture}
\label{fig_Method}
\end{figure}

In the context of the proposed MDMT model, we propose a novel framework designed to leverage shared and task-specific features across multiple datasets, thereby enhancing the model's ability to generalize and perform distinct tasks simultaneously.
This approach not only capitalizes on the inherent diversity and complementarity of multi-source data, but also introduces a flexible architecture that adapts to the unique requirements of each task through dedicated heads and a unified backbone for feature extraction.
In Figure~\ref{fig_Method}, we report an overview of the proposed MDMT model.

Given two datasets, $D^{\tau_1}$ and $D^{\tau_2}$, each referring to a task $\tau_1$ and $\tau_2$, respectively, a single batch $X$ is formed by randomly selecting elements such that $X_i \in D^{\tau_1} \cup D^{\tau_2}$ for any element $X_i$ in $X$.
A shared backbone feature extraction network, denoted as $f^s$, processes $X$ to yield a set of shared features $H$.
This shared representation is then fed into two task-specific fully connected network heads, $f^{\tau_1}$ and $f^{\tau_2}$, for tasks $\tau_1$ and $\tau_2$, respectively, producing outputs $O^{\tau_1}$ and $O^{\tau_2}$.
The model employs an indicator function $\mathcal{I}(X_i, \tau_j)$ to determine the association of a sample $X_i$ from batch $X$ to a task $\tau_j$, crucial for task-specific loss computations.
The losses for the tasks $\tau_1$ and $\tau_2$, represented as $\mathcal{L}^{\tau_1}$ and $\mathcal{L}^{\tau_2}$, respectively, are computed using $O^{\tau_1}$, $Y^{\tau_1}$ and $O^{\tau_2}$, $Y^{\tau_2}$, where $Y^{\tau_1}$ and $Y^{\tau_2}$ denote the true labels of $X$ for the respective tasks.

In the following, we delve deeper into the specifics of each component within the network architecture, encompassing the shared backbone, the task-specific heads and the MDMT learning mechanism.

In our model, the shared backbone consists of CNN layers that act as a common trainable module for the flow of information regardless of the tasks.
The output of this shared network is represented by the function $H = f^s(X; \theta^s)$, where $X$ denotes the input batch from the combined datasets $D^{\tau_1}$ and $D^{\tau_2}$, and $\theta^s$ symbolizes the trainable parameters of the CNN backbone.
This unified feature extraction mechanism is crucial for capturing generalized representations that are beneficial across multiple tasks.

In our architecture, we enhance the shared CNN backbone by appending two task-specific heads of fully connected layers, which identify distinct task-specific modules for $D^{\tau_1}$ and $D^{\tau_2}$ and their respective tasks. 
The output vector for each task-specific branch is defined as:
\begin{equation}
 O^{\tau_j} = f^{\tau_j}(H, \theta^{\tau_j}) , \quad j \in \{1,2\}
\end{equation}
Where $H$ is the output from the shared backbone and $\theta^{\tau_j}$ are the parameters of the task-specific fully connected networks, ensuring a dedicated head for processing the features according to the specific requirements of each task.

The uniqueness of our approach lies in the integration of the two loss functions $ \mathcal{L}^{\tau_1} $ and $ \mathcal{L}^{\tau_2} $ through the adoption of an indicator function, inspired by Lee et al.'s approach to handling censored data in overall survival analysis~\cite{lee2018deephit}.
This adaptation is crucial given the context where, within a mixed batch, no samples bear labels for both tasks concurrently, making it impractical to directly apply a loss function across different datasets.
Hence, the proposed total loss function, $\mathcal{L}$, is crafted as:
\begin{equation}
 \mathcal{L} = \sum_{i = 1}^{|X|} \mathcal{I}(X_i, \tau_1) \cdot \mathcal{L}^{\tau_1}(O^{\tau_1}_i, Y^{\tau_1}_i) + \mathcal{I}(X_i, \tau_2) \cdot \mathcal{L}^{\tau_2}(O^{\tau_1}_i, Y^{\tau_1}_i)
\label{eq:ltot}
\end{equation}
where $|X|$ is the number of instances in $X$.

The indicator function $\mathcal{I}(X_i,\tau_j)$, plays a pivotal role in the MDMT model by allowing the integration of multi-source data into a unified framework, especially when computing task-specific losses.
This function is designed to identify the association of a given sample with its specific task, which is crucial for models that handle data from multiple datasets to perform multiple task optimization simultaneously.
It can be defined as:
\[
\mathcal{I}(X_i, \tau_j) = 
\begin{cases} 
1 & \text{if sample } X_i \text{ is associated with task } \tau_j \\
0 & \text{otherwise.}
\end{cases}
\]
Since the batch $X$ contains elements from both datasets $D^{\tau_1}$ and $D^{\tau_2}$, and each dataset may be associated with different tasks, $\mathcal{I}(X_i,\tau_j)$ helps to selectively compute the loss for each task by including only those samples related to the task in question.
The use of an indicator function in this manner adds flexibility and scalability to the model. It allows for the seamless addition of new tasks or datasets by simply adjusting the indicator function to include new elements in the loss computation process without needing significant architectural changes.
In the proposed MDMT model, the training process is designed to be end-to-end, indicating that all model components are trained simultaneously during the training phase.
This approach ensures that the model learns to optimize its parameters in a cohesive manner, leveraging gradients derived from the total loss to update each component of the architecture.

For the two task-specific loss functions, we selected $\mathcal{L}^{\tau_1}$ and $\mathcal{L}^{\tau_2}$ considering each task characteristics and labeling schemes, as detailed in Section~\ref{sec:experimental_setup}.
Given that $\tau_1$, for the severity group prediction, and $\tau_2$, for the severity assessment, are both classification tasks, we selected the Cross-Entropy Loss, defined as follows:
\begin{equation}
 \mathcal{L}^{\tau_j}(O^{\tau_j}_i, Y^{\tau_j}_i) = -\sum_{k=1}^{c^{\tau_j}} Y^{\tau_j}_{i,k} \log(O^{\tau_j}_{i,k})
\label{eq:BCE}
\end{equation}
where $Y^{\tau_j}_i$ is the real label, $O^{\tau_j}_i$ the predicted probabilities for the sample $X_i$ and $c^{\tau_j}$ is the number of class labels for the considered task $\tau_j$.
These loss functions are tailored to the distinct characteristics of each task within the model, allowing for precise adjustments to the model parameters based on the nature of the tasks at hand.
The ability to compute task-specific losses accurately is crucial for the model to balance learning across tasks, preventing the dominance of any single-task over the others and promoting a synergistic improvement in performance.

\section{Experimental Setup} \label{sec:experimental_setup}

This section presents our evaluation framework for the MDMT model against conventional approaches, detailing the datasets, preprocessing steps and experimental configurations.

\subsection{Datasets and Tasks}
Our study leverages two distinct and publicly available CXR datasets, namely AIforCOVID~\cite{soda2021aiforcovid} and BRIXIA~\cite{signoroni2021severity}, each assigned to a distinct task: severity prognosis and severity assessment, respectively.
These datasets were specifically chosen for their complementary characteristics, enabling a comprehensive assessment of the model's ability to generalize across different tasks and data distributions.

The AIforCOVID dataset $D^{\tau_1}$ encompasses 1586 CXR examinations from COVID-19 positive adult patients, confirmed via RT-PCR tests, from six different centers.
This dataset has been compiled into three releases, each expanding the dataset's size: the initial release included 820 patients, followed by an additional 284 patients, and concluded with an additional 482 patients.
Patients are classified into two categories based on treatment outcomes: mild (home isolation or hospitalization without ventilatory support) and severe (requiring non-invasive ventilation, ICU admission, or resulting in death), making the task a binary classification problem, with the target variable $Y^{\tau_1}_i \in \{0, 1\}$ representing these outcomes.

The BRIXIA dataset $D^{\tau_2}$ consists of 4707 CXR images from COVID-19 positive patients treated in sub-intensive and intensive care units, obtained from various machine manufacturers and varied acquisition parameters.
A noteworthy feature of this dataset is the Brixia score, which grades lung opacity across six regions on a scale of $0$ to $3$.
To simplify the scoring system, we sum the regional scores into a global score $G_i$ for each image $X_i$.
This score is then categorized into four severity levels through a threshold function:
\begin{equation}
Y^{\tau_2}_i = 
\begin{cases}
G_i < 5, & \text{Category 0} \\
5 \leq G_i < 9, & \text{Category 1} \\
9 \leq G_i < 14, & \text{Category 2} \\
G_i \geq 14, & \text{Category 3}
\end{cases}
\end{equation}
This categorization yields a multi-class target variable $Y^{\tau_2}_i \in \{ 0, 1, 2, 3\}$, facilitating the model's task of predicting the severity of lung involvement in COVID-19 patients.

To ensure the data from both AIforCOVID and BRIXIA datasets are coherent and prepared for our model, we employed a preprocessing pipeline that includes lung segmentation via a U-NET model, bounding-box extraction, resizing and normalization.
For ease of reproducibility, these steps are the same used in~\cite{guarrasi2022optimized,guarrasi2023multi,guarrasi2021multi,guarrasi2022pareto}.

\subsection{Experimental Configurations}

To comprehensively validate the efficacy of our MDMT model, we designed three distinct experimental configurations, aimed at demonstrating the model's superiority over traditional STL and fine-tuning approaches.

We establish a baseline by training the CNN backbone on each dataset and task separately. This step serves as a foundation for comparison, highlighting the potential enhancements brought by MTL. The experiments are denoted as $\mathit{STL}^{\tau_1}$ and $\mathit{STL}^{\tau_2}$ for $D^{\tau_1}$ and $D^{\tau_2}$ datasets, respectively.
Each backbone utilized in this phase is pre-trained on the ImageNet dataset~\cite{ke2021chextransfer}.

As a comparative benchmark, we explore a conventional fine-tuning strategy where the model pre-trained on the BRIXIA dataset $\mathit{STL}^{\tau_2}$ is fine-tuned for the AIforCOVID task $D^{\tau_1}$. This experiment, denoted as $\mathit{FT}$, investigates the effectiveness of transferring learned features from one task to another, contrasting it against the integrated learning approach of the MDMT model.

The proposed MDMT Learning model employs as pre-training the average of the weights from the baseline models $\mathit{STL}^{\tau_1}$ and $\mathit{STL}^{\tau_2}$. This experiment, denoted $\mathit{MDMT}$, allows simultaneous training on both tasks, leveraging shared and task-specific features, as detailed in Section~\ref{section:method}. 
 
To evaluate the robustness of the proposed methodology, all the aforementioned experiments are conducted on 18 CNN backbones across 6 major architectural families: DenseNet, EfficientNet, GoogLeNet, MobileNet, ResNet and ShuffleNet.
For all experiments and networks, uniform training configurations were applied. Optimization was carried out using the Adam optimizer, with an initial learning rate of 0.001, momentum of 0.9, and weight decay of 0.0001. Training included a 40-epoch warm-up period and was limited to a maximum of 300 epochs, incorporating early stopping after 40 epochs to prevent overfitting. All training sessions were conducted on four NVIDIA TESLA A100 GPUs, utilizing a batch size of 128.

To ensure the reliability and generalizability of our findings, we employ both 5-fold stratified cross-validation (CV) and leave-one-center-out (LOCO) validation strategies.
By evaluating the Accuracy (ACC), F1-score (F1) and G-mean (GM), we perform a comprehensive evaluation of the model's performance across different partitions of the data, highlighting its ability to generalize to unseen data and mitigate bias towards specific data distributions.

\section{Results and Discussions}

To evaluate the effectiveness of our proposed MDMT method, we compared our MDMT framework with each experimental configuration, validation strategy and CNN backbone employed, as shown in Table~\ref{table:results_avg} and in more detail in Table~\ref{table:results_m1}. We opted to discuss the results obtained on $\tau_1$ task, given its unique relationship with the task $\tau_2$.
The task $\tau_2$ not only temporally precedes $\tau_1$ but also possesses strong predictive power for prognostic prediction, therefore incorporating $\tau_2$ can lead to performance improvement on $\tau_1$.

To provide a comprehensive overview of our results, Table~\ref{table:results_avg} displays the average performance metrics across all experimental configurations and CNN architectures tested in our study. Our analysis reveals a clear superiority of the MDMT approach over the conventional methodologies, including both the STL and fine-tuning strategies.
In Table~\ref{table:results_m1}, all the performances for all the shared backbones are reported. These results not only confirm MDMT's superior average performance, but also demonstrate that our method achieves the highest absolute performance over backbone architectures, experimental scenarios and validation strategies.

\begin{table}[t]
\scriptsize
\centering
\resizebox{0.75\linewidth}{!}{
\begin{tabular}{l|c|c|c|c|c|c}
\hline
\multirow{2}{*}{\textbf{Experiment}} & \multicolumn{3}{c|}{\textbf{CV}} & \multicolumn{3}{c}
{\textbf{LOCO}} \\
\cline{2-7}
& \textbf{ACC} & \textbf{F1} & \textbf{GM} & \textbf{ACC} & \textbf{F1} & \textbf{GM} \\
\hline
$\mathit{STL}^{\tau_1}$ & 66.0 & 63.9 & 65.9 & 61.2 & 57.3 & 60.1 \\
$\mathit{FT}$ & 67.3 & 63.1 & 66.9 & 64.8 & 58.9 & 63.8 \\
$\mathit{MDMT}$ & \textbf{68.6} & \textbf{66.6} & \textbf{68.5} & \textbf{65.7} & \textbf{64.3} & \textbf{66.0} \\
\hline
\end{tabular}}
\caption{Average model performance metrics for task $\tau_1$, calculated across different backbone architectures.}
\label{table:results_avg}
\end{table}

\begin{table}[t]
\scriptsize
\centering
\resizebox{0.75\linewidth}{!}{
\begin{tabular}{l|c|c|c|c|c|c|c}
\hline
\multirow{2}{*}{\textbf{Statistic}} & \multirow{2}{*}{\textbf{Test}} & \multicolumn{3}{c|}{\textbf{CV}} & \multicolumn{3}{c}
{\textbf{LOCO}} \\
\cline{3-8}
 & & \textbf{ACC} & \textbf{F1} & \textbf{GM} & \textbf{ACC} & \textbf{F1} & \textbf{GM} \\
\hline
\multirow{2}{*}{$\mu$} & 
$\mathit{MDMT}$ vs $\mathit{STL}^{\tau_1}$ & $***$ & $***$ & $***$ & $***$ & $***$ & $***$ \\ \cline{2-8} &
$\mathit{MDMT}$ vs $\mathit{FT}$ & $**$ & $***$ & $**$ & & & \\ 
\hline
\multirow{2}{*}{$\sigma$} 
& $\mathit{MDMT}$ vs $\mathit{STL}^{\tau_1}$ & $*$ & $*$ & $*$ & $*$ & & $*$ \\ \cline{2-8}
& $\mathit{MDMT}$ vs $\mathit{FT}$ & & & & $*$ & & $*$ \\ 
\hline
\end{tabular}}
\caption{Statistical comparison of $\mathit{MDMT}$ vs $\mathit{STL}^{\tau_1}$ and $\mathit{MDMT}$ vs $\mathit{FT}$. Significance levels are marked with $*$ for a $p$-value $< 0.05$, $**$ for a $p$-value $< 0.01$ and $***$ for a $p$-value $< 0.001$.}
\label{table:results_stat}
\end{table}

To rigorously evaluate the effectiveness of the proposed MDMT method, we employed a single-tail $t$-test for statistical comparisons across the various experimental configurations. This analysis involved assessing both the average performance metrics $\mu$ and their standard deviations $\sigma$ across test folds. Our goal was to determine whether there was a statistically significant increase in $\mu$ and a decrease in $\sigma$ for $\mathit{MDMT}$ in comparison to the $\mathit{STL}^{\tau_1}$ and the $\mathit{FT}$ settings. Notably, an increase in $\mu$ indicates a significant increase in performance and a reduction of $\sigma$ indicates enhanced model robustness. The findings, detailed in Table~\ref{table:results_stat}, show the statistical significance achieved by MDMT.

Analyzing the results in CV, we observed that the $\mathit{MDMT}$ method consistently outperformed the $\mathit{STL}^{\tau_1}$
and the $\mathit{FT}$ approaches in terms of average performance metrics across the CNN architectures, further validated through statistical comparisons.
When compared to $\mathit{STL}^{\tau_1}$, $\mathit{MDMT}$ not only demonstrates superior performance across all metrics but also shows a reduction in standard deviation, indicating both higher performance and increased robustness. In contrast, while $\mathit{MDMT}$ exhibits statistically significant performance enhancements over $\mathit{FT}$, the improvement in robustness is not statistically significant. To clarify, we conducted additional tests on $\sigma$, confirming that $\mathit{MDMT}$'s robustness is statistically comparable to that of the fine-tuning strategies.

In the LOCO validation scenario, $\mathit{MDMT}$ demonstrated superior performance relative to both the $\mathit{STL}^{\tau_1}$ and the $\mathit{FT}$ methods. This performance superiority underscores the robustness of $\mathit{MDMT}$ across various backbone architectures and its enhanced ability to generalize across different clinical settings. The statistical analysis highlights a significant improvement in performance metrics for $\mathit{MDMT}$ compared to $\mathit{STL}^{\tau_1}$. 
Regarding the standard deviation, the reduction in variability for ACC and F1 metrics suggests increased reliability of $\mathit{MDMT}$ over the compared methods. To address any potential concerns about non-significant differences, we conducted one-tail $t$-tests, which confirmed that $\mathit{MDMT}$'s performance is not statistically inferior to that of the $\mathit{FT}$ approach.

For brevity, we presented results only from the final release, though all experiments were conducted across all three releases of the AIforCOVID dataset, yielding consistent results and conclusions.
Furthermore, all experiments and analyses were similarly conducted on $\tau_2$, showing no significant change in performance with respect to the $\mathit{STL}^{\tau_2}$ and the corresponding $\mathit{FT}$, underscoring that while $\tau_2$ benefits $\tau_1$, the converse is not necessarily true.

\section{Conclusion}
In this study, we introduced a MDMT approach aimed at enhancing the generalization capabilities of deep neural networks for COVID-19 prognosis prediction. By concurrently learning from two distinct datasets for different tasks, our method not only showcases higher performance across various experimental benchmarks but also establishes itself as a robust solution within the biomedical field for the integration of diverse datasets into a unified training framework, leading to a multi-perspective analysis of the disease. 
Future directions for this research include expanding its application to other domains, together with the adoption of explainable AI techniques~\cite{guarrasi2022multimodal} to understand the importance of each task and the integration of multi-modal data sources aimed to achieve a holistic understanding of disease aspects, thereby leading to more precise, trustworthy and complete systems. 

\section*{Acknowledgements}

Filippo Ruffini is a PhD student enrolled in the National PhD in Artificial Intelligence, XXXVIII cycle, course on Health and life sciences, organized by Università Campus Bio-Medico di Roma. 
This work was partially supported by: i) the Italian Ministry of Foreign Affairs and International Cooperation, grant number PGR01156, ii) PNRR MUR project PE0000013-FAIR, iii) PNRR – DM 118/2023.
Resources are provided by the National Academic Infrastructure for Supercomputing in Sweden (NAISS) and the Swedish National Infrastructure for Computing (SNIC) at Alvis @ C3SE, partially funded by the Swedish Research Council through grant agreements no. 2022-06725 and no. 2018-05973.

\bibliographystyle{splncs04}
\bibliography{main}

\newpage
\beginsupplement

\textbf{\large Supplementary Material}

\begin{table}[!h]
\scriptsize
\centering
\resizebox{\linewidth}{!}{
\begin{tabular}{l|l|c|c|c|c|c|c}
\hline
\multirow{2}{*}{\textbf{Backbone}} & \multirow{2}{*}{\textbf{Experiment}} & \multicolumn{3}{c|}{\textbf{CV}} & \multicolumn{3}{c}
{\textbf{LOCO}} \\
\cline{3-8}
 & & \textbf{ACC} & \textbf{F1} & \textbf{GM} & \textbf{ACC} & \textbf{F1} & \textbf{GM} \\
\hline
\multirow{3}{*}{\textbf{DenseNet-121}} & $\mathit{STL}^{\tau_1}$ & 66.1(4.3) & 65.0(5.2) & 66.2(3.8) & 61.2(7.5) & 53.0(21.3) & 59.2(4.2) \\
& $\mathit{FT}$ & 67.3(2.2) & 63.1(4.1) & 66.9(2.1) & 64.2(6.0) & 60.9(9.6) & 63.1(3.1)\\
& $\mathit{MDMT}$ & \textbf{69.9(2.8)} & \textbf{69.1(3.4)} & \textbf{70.0(2.5)} & \textbf{65.6(4.2)} & \textbf{65.5(7.0)} & \textbf{66.4(3.5)} \\
\hline
\multirow{3}{*}{\textbf{DenseNet-161}} & $\mathit{STL}^{\tau_1}$ & 67.4(2.5) & 65.8(3.6) & 67.4(2.3) & 62.1(7.4) & 59.5(14.5) & 60.3(3.9) \\
& $\mathit{FT}$ & 65.8(3.1) & 60.7(3.7) & 65.3(2.8)& 62.9(8.3) & 52.6(22.1) & 62.0(5.4)\\
& $\mathit{MDMT}$ & \textbf{68.7(2.8)} & \textbf{67.0(2.9)} & \textbf{68.6(2.4)} & \textbf{68.5(6.6)} & \textbf{64.8(10.7)} & \textbf{67.4(4.8)} \\
\hline
\multirow{3}{*}{\textbf{DenseNet-169}} & $\mathit{STL}^{\tau_1}$ & 67.1(2.4) & 65.5(3.2) & 67.1(2.2) & 57.9(9.1) & 51.0(26.2) & 57.2(4.8)\\
& $\mathit{FT}$ & 67.6(3.6) & 64.4(3.9) & 67.3(3.2)& 64.8(6.3) & 62.6(10.1) & 62.9(5.5) \\
& $\mathit{MDMT}$ & \textbf{68.5(2.5)} & \textbf{66.8(2.6)} & \textbf{68.4(2.2)} & \textbf{65.9(6.3)} & \textbf{62.9(11.5)} & \textbf{65.8(5.5)} \\
\hline
\multirow{3}{*}{\textbf{DenseNet-201}} & $\mathit{STL}^{\tau_1}$ & 66.9(2.2) & 64.56(3.3) & 66.7(2.0) & 63.5(6.8) & 61.5(13.2) & 61.3(2.3)\\ 
& $\mathit{FT}$ & 66.3(1.7) & 61.0(3.2) & 65.7(1.5) & 66.0(6.5) & 62.6(10.7) & 65.0(5.0) \\
& $\mathit{MDMT}$ & \textbf{69.2(2.2)} & \textbf{66.5(3.4)} & \textbf{69.0(2.0)} & \textbf{69.5(5.4)} & \textbf{68.9(7.3)} & \textbf{68.8(4.2)} \\
\hline
\multirow{3}{*}{\textbf{EfficientNet-b0}} & $\mathit{STL}^{\tau_1}$ & 65.5(1.8) & 62.8(5.5) & 65.3(2.0) & 59.5(8.1) & 56.7(16.4) & 59.1(5.3)\\
& $\mathit{FT}$ & 68.5(0.9) & 66.2(1.3) & 68.4(0.8) & \textbf{68.9(5.1)} & \textbf{67.5(8.2)} & \textbf{68.3(4.0)} \\
& $\mathit{MDMT}$ & \textbf{70.1(1.5)} & \textbf{68.7(2.2)} & \textbf{70.1(1.2)} & 66.3(6.4) & 65.5(6.5) & 67.4(5.3) \\
\hline
\multirow{3}{*}{\textbf{EfficientNet-b1-p}} & $\mathit{STL}^{\tau_1}$ & 66.0(3.2) & 64.2(2.8) & 66.0(2.9) & 59.1(7.6) & 60.2(12.1) & 57.9(4.1)\\
& $\mathit{FT}$ & \textbf{69.0(1.8)} & \textbf{67.0(1.9)} & \textbf{68.9(1.4)} & \textbf{65.7(8.8)} & \textbf{60.5(17.7)} & \textbf{65.0(6.0)} \\
& $\mathit{MDMT}$ & 68.1(0.6) & 66.4(2.4) & 68.0(0.6) & 59.3(12.2) & 55.6(16.8) & 61.8(6.3) \\
\hline
\multirow{3}{*}{\textbf{EfficientNet-es}} & $\mathit{STL}^{\tau_1}$ & 66.0(3.8) & 65.6(5.2) & 66.2(3.5) & 54.8(14.1) & 43.9(23.2) & 57.8(7.6)\\
& $\mathit{FT}$ & \textbf{68.1(1.7)} & \textbf{64.3(2.5)} &\textbf{ 67.8(1.6)} & 65.0(5.9) & 61.8(10.8) & 63.6(4.1) \\
& $\mathit{MDMT}$ & 67.9(3.0) & 63.7(4.5) & 67.5(2.8) & \textbf{66.9(4.9)} & \textbf{65.5(8.5)} & \textbf{66.1(2.4) }\\
\hline
\multirow{3}{*}{\textbf{EfficientNet-es-p}} & $\mathit{STL}^{\tau_1}$ & 64.8(3.1) & 62.0(5.6) & 64.6(3.1) & 60.3(9.1) & 56.4(19.6) & 60.2(5.2)\\
& $\mathit{FT}$ & 67.5(2.5) & 61.1(5.1) & 66.8(2.5) & 64.3(10.7) & 58.0(21.7) & 63.5(8.0) \\
& $\mathit{MDMT}$ & \textbf{68.5(2.3)} & \textbf{66.0(2.2)} & \textbf{68.3(1.9)} & \textbf{67.1(4.5)} & \textbf{67.4(5.4)} & \textbf{66.7(3.6)} \\
\hline
\multirow{3}{*}{\textbf{EfficientNet-lite}} & $\mathit{STL}^{\tau_1}$ & 63.6(2.5) & 60.1(5.2) & 63.4(2.5) & 59.6(8.0) & 58.7(8.0) & 61.0(6.0)\\
& $\mathit{FT}$ & 68.0(1.7) & 64.5(2.3) & 67.7(1.6) & \textbf{66.1(6.1)} & \textbf{62.5(11.4)} & \textbf{65.0(3.5)}\\
& $\mathit{MDMT}$ & \textbf{68.4(1.6)} & \textbf{67.2(3.1)} & \textbf{68.4(1.5)} & 60.3(5.7) & 59.2(5.7) & 62.3(3.8) \\
\hline
\multirow{3}{*}{\textbf{GoogLeNet}} & $\mathit{STL}^{\tau_1}$ & 65.8(4.1) & 63.4(4.7) & 65.6(3.7) & 62.0(9.9) & 55.5(23.2) & 60.1(4.8)\\
& $\mathit{FT}$ & 65.8(2.7) & 60.9(4.0) & 65.3(2.5) & 64.3(5.3) & 58.0(13.1) & 64.0(4.5) \\
& $\mathit{MDMT}$ & \textbf{68.1(3.1)} & \textbf{66.4(3.6)} & \textbf{68.0(2.8)} & \textbf{66.8(6.0)} & \textbf{64.4(11.4)} & \textbf{66.6(5.1)} \\
\hline
\multirow{3}{*}{\textbf{MobileNet}} & $\mathit{STL}^{\tau_1}$ & 66.6(3.1) & 63.7(3.6) & 66.3(2.8) & 62.5(4.8) & 61.6(13.5) & 59.7(3.7)\\
& $\mathit{FT}$ & 68.9(1.5) & 65.8(1.3) & 68.6(1.3) & \textbf{67.5(5.6)} & 66.5(8.6) & \textbf{66.3(4.4)} \\
& $\mathit{MDMT}$ & \textbf{69.1(1.7)} & \textbf{67.2(2.0)} & \textbf{69.0(1.4)} & \textbf{67.5(7.2)} & \textbf{66.3(9.4)} & 66.7(6.3) \\
\hline
\multirow{3}{*}{\textbf{ResNet-18}} & $\mathit{STL}^{\tau_1}$ & 66.6(3.1) & 63.7(3.6) & 66.3(2.8) & 64.4(7.4) & 62.4(12.6) & 61.6(3.6)\\
& $\mathit{FT}$ & 66.7(3.0) & 62.3(3.9) & 66.3(2.7) & 64.5(5.6) & 56.0(15.4) & 62.9(3.6) \\
& $\mathit{MDMT}$ & \textbf{68.5(2.7)} & \textbf{67.7(3.4)} & \textbf{68.6(2.5)} & \textbf{67.1(6.9)} & \textbf{66.6 (9.4)} & \textbf{67.2(5.2)} \\
\hline
\multirow{3}{*}{\textbf{ResNet-34}} & $\mathit{STL}^{\tau_1}$ & 67.0(2.1) & 64.4(3.9) & 66.8(2.1) & 61.2(7.8) & 49.8(21.7) & 58.5(3.4)\\
& $\mathit{FT}$ & 66.9(2.5) & 62.5(3.9) & 66.5(2.3) & \textbf{64.2(6.5)} & 56.6(18.8) & 63.4(4.4) \\
& $\mathit{MDMT}$ & \textbf{69.3(2.7)} & \textbf{66.8(1.6)} & \textbf{69.1(2.2)} & 63.3(9.5) & \textbf{60.4(10.8)} & \textbf{64.9(6.8)}\\
\hline
\multirow{3}{*}{\textbf{ResNet-50}} & $\mathit{STL}^{\tau_1}$ & 67.2(4.2) & 65.2(3.7) & \textbf{67.1(4.0)} & \textbf{65.2(8.7)} & \textbf{67.1(9.6)} & 63.2(5.7)\\
& $\mathit{FT}$ & 66.4(2.2) & 61.7(2.4) & 65.9(1.9) & 63.2(9.6) & 53.7(25.8) & 62.1(6.8) \\
& $\mathit{MDMT}$ & \textbf{67.2(2.2)} & \textbf{65.6(1.7)} & \textbf{67.1(1.7)} & 64.1(3.9) & 61.6(8.4) & \textbf{65.6(3.9)} \\
\hline
\multirow{3}{*}{\textbf{ResNet-101}} & $\mathit{STL}^{\tau_1}$ & 66.2(1.5) & 63.6(2.2) & 66.0(1.0) & 61.7(9.2) & 54.6(20.6) & 59.2(4.3) \\
& $\mathit{FT}$ & 67.1(2.6) & 60.8(3.4) & 66.4(2.3) & 62.1(8.4) & 49.1(23.5) & 60.4(4.9) \\
& $\mathit{MDMT}$ & \textbf{69.5(1.9)} & \textbf{68.0(2.6)} & \textbf{69.5(1.7)} & \textbf{67.1(9.9)} & \textbf{65.6(12.1)} & \textbf{66.6(7.0)} \\
\hline
\multirow{3}{*}{\textbf{ShuffleNet-v2-x0}} & $\mathit{STL}^{\tau_1}$ & 65.9(2.0) & 63.3(3.4) & 65.7(2.0) & 61.9(8.0) & 61.8(12.1) & 61.34(4.1)\\
& $\mathit{FT}$ & \textbf{68.2(1.5)} & 64.0(3.1) & \textbf{67.8(1.5)} & 64.9(5.9) & 62.7(9.5) & 64.0(3.4) \\
& $\mathit{MDMT}$ &67.8(1.2) & \textbf{64.7(1.6)} & 67.5(1.1) & \textbf{66.2(6.0)}&	\textbf{66.0(7.9)}& \textbf{65.8(3.7)} \\
\hline
\multirow{3}{*}{\textbf{ShuffleNet-v2-x1}} & $\mathit{STL}^{\tau_1}$ & 64.1(3.3) & 62.4(5.0) & 64.0(3.0) & 62.4(6.8) & 63.1(12.6) & 62.9(3.8)\\
& $\mathit{FT}$ & 67.8(1.5) & 63.2(1.5) & 67.3(1.3) & 65.2(8.9) & 56.7(21.9) & 64.5(6.7) \\
& $\mathit{MDMT}$ & \textbf{68.8(2.1)} & \textbf{66.9(3.8)} & \textbf{68.7(2.1)} & \textbf{65.5(6.0)}&	\textbf{66.0(8.2)}& \textbf{65.8(4.5)}\\
\hline
\multirow{3}{*}{\textbf{Wide-ResNet50-2}} & $\mathit{STL}^{\tau_1}$ & 65.6(2.4) & 65.2(2.4) & 65.8(2.0) & 61.6(6.5) & 54.8(14.9) & 61.2(3.6)\\
& $\mathit{FT}$ & 66.0(2.2) & 61.4(2.2) & 65.6(1.9) & 63.1(9.8) & 52.0(26.6) & 61.8(6.7) \\
& $\mathit{MDMT}$ & \textbf{66.3(1.9)} & \textbf{64.9(1.5)} & \textbf{66.3(1.6)} & \textbf{66.1(4.7)} & \textbf{66.0(8.4)} & \textbf{65.9(3.6)} \\
\hline
\end{tabular}}
\caption{Model performance metrics on task $\tau_1$, including mean and standard deviation across folds, for the experiments and validation approaches.}
\label{table:results_m1}
\end{table}

\end{document}